# "SHADOWING" OF THE ELECTROMAGNETIC FIELD OF A RELATIVISTIC ELECTRON[*]

G. NAUMENKO

*Nuclear Physics Institute of Tomsk Polytechnic University, Lenina str. 2a*
*Tomsk, 634050, Russia*

X. ARTRU

*Institut de Physique Nucl.eaire de Lyon, Université de Lyon, CNRS- IN2P3 and*
*Université Lyon 1*
*69622 Villeurbanne, France*

A. POTYLITSYN

*Tomsk Polytechnic University, Lenina str. 2*
*Tomsk, 634050, Russia*

YU. POPOV

*Tomsk Polytechnic University, Lenina str. 2*
*Tomsk, 634050, Russia*

L. SUKHIKH

*Tomsk Polytechnic University, Lenina str. 2*
*Tomsk, 634050, Russia*

In coherent radiation sources (diffraction radiation, Smith-Purcell effect, etc.) based on relativistic electrons passing by a material "radiator", the electron self-field is partly shadowed after each part of the radiator over a distance of the order of the formation length $\gamma^2\lambda$. This effect has been investigated on coherent diffraction radiation (DR) by electron bunches. An absorbing half-plane screen was placed at various distances $L$ before a standard DR source (inclined half-plane mirror). The DR intensity was reduced when the screen was at small $L$ and on the same side as the mirror. No reduction was observed when the screen was on the opposite side. The shadowing effect can significantly reduce the total energy radiated in a long radiator.

---

[*] This work was partly supported by the warrant-order 1.226.08 of the Ministry of Education and Science of the Russian Federation..

## 1. introduction

Although the radiation emitted by relativistic electrons passing through or near material targets can be calculated (in principle) from the macroscopic Maxwell equations, phenomenological concepts, like *formation zone* and *equivalent photons*, are useful for the intuitive understanding of the main features. The formation zone is a region of length $l_f \sim \gamma^2 \lambda$ where the Coulomb field of the fast electron interfere with the forward radiation field (in this paper we assume that the electron is ultrarelativistic : $\gamma \gg 1$). The interference is destructive, therefore the electron is said to be "half-naked" [1]. The concept of half-naked electron, associated with the Equivalent Photon Method, can explain in a simple way the suppression of Optical Transition Radiation in a Wartski interferometer when the distance between the two foils is small compared to $l_f$, or the Landau-Pomeranchuk-Migdal effect in Bremsstrahlung.

In this paper, it will be shown both experimentally and theoretically that behind a Diffraction Radiation (DR) target there is a *shadow* region where the electromagnetic field is much smaller than the normal Coulomb field of the particle, and radiation by a second DR target is inhibited. This region is similar to the usual formation zone but only on one side of the particle trajectory.

The shadowing concept will be introduced more precisely in the next section. Its consequences on the Smith-Purcell Radiation will be pointed out. Section 3 will give the principle of the experiment conducted at the Tomsk Nuclear Physics Institute. Theoretical calculations are presented in Section 4, with details in Appendix A. The experimental setup and method are presented in Section 5, the results in Section 6. The latter, as well as further questions, are discussed in Section 7.

## 2. Shadowing effect from different points of view

Here we present different points of view about the shadow effect. Let us consider a fast electron moving near two targets like in Fig.1.

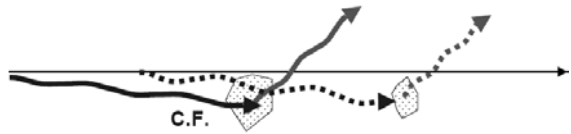

Fig.1. Shadowing of the target by another target

The Coulomb field (C.F.) is considered as a beam of quasi-real photon. Scattering of this field by the first target gives Diffraction Radiation. The second target is in the shadow of the first one, therefore emits almost no radiation. The Coulomb field is gradually "repaired", and the shadow disappears, during the formation zone of length $l_f \sim \gamma b \sim \gamma^2 \lambda$, where $b$ (impact parameter) is the distance between the trajectory and the target (typically, $\lambda \sim \gamma/b$).

Similarly, after the scattering of a charged particle at large angle, there is a region where the Coulomb field is partly missing. The term "half-naked electron" [1] has been introduced to describe this effect, in the framework of quantum electrodynamics.

A geometrical point of view is given in [13], where Figs.1.1 and 2.4 depict the causal restoration of the Coulomb field after the electron scattering. In Fig.2 we adapted this picture to the case of a particle passing through a narrow hole, and in Fig.3 to a particle passing near a half-screen, in the Diffraction Radiation geometry.

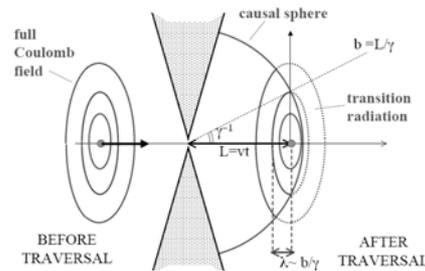

Fig. 2. Field of a particle passing through a hole (analogous to Fig.1.1 of [13]).

In Fig.2, the Coulomb field is removed in all directions about the trajectory. In Fig.3, only one side is removed. This asymmetry has been tested by the experiment presented below.

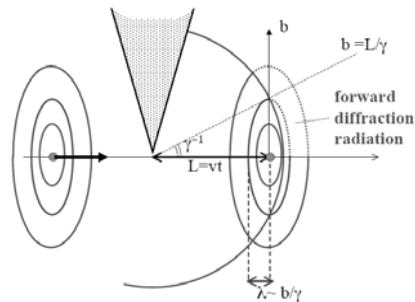

Fig. 3. Shadowing by a half-screen.

Another point of view is shown in Fig.4. The Coulomb field (C.F.) and the forward diffracted radiation (FDR) of the first target interfere destructively. We can also say that the FDR from the second target exists, but it interferes destructively with the rescattered part of the FDR from the first target. Shadow effect is then a rescattering effect, like the dynamical effect in PXR.

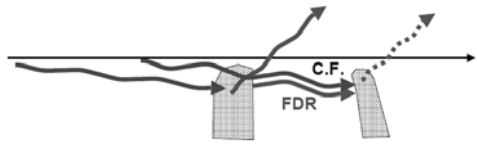

Fig. 4. Shadow effect as a re-scattering effect.

This point of view was already formulated by B.M. Bolotovskii [12], who considered a charged particle passing near a screen, as shown in Fig.5.

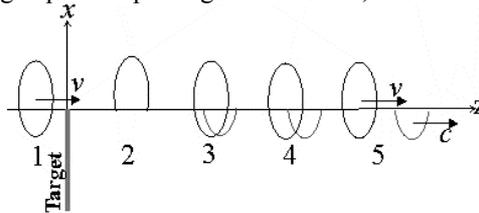

Fig. 5. Presentation of the radiation formation length effect by B.M. BolotovskiI in [12].

The ellipse 1 represents the field of the particle. It hits the half-plane screen and induces current in it, which in turn emit Diffraction Radiation. "It results the peculiar features of the radiation field. The radiation field is such that close to the screen it kills part of the particle field". At positions 3, 4, 5 we see the separation of the radiation field and the Coulomb field, due to their different velocities.

## 2.1. *Consequences of the shadow effect in Smith-Purcell radiation.*

An example of Smith-Purcell (SP) radiator is the periodic set of foils shown in Fig.6. The mere addition of the Diffraction Radiation amplitudes from the different foils neglects the shadow effect, therefore over-estimate the SP intensity.

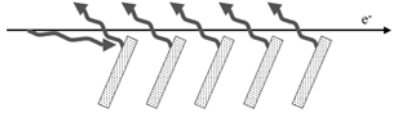

Fig. 6. Shadowing in a periodical target.

For a given impact parameter $b$, the foil spacing which gives the maximum SP energy per unit length, $dW/dz$, is of the order of $l_f \sim \gamma b$: for smaller spacing shadowing is important, for larger spacing there are too few foils. The optimal value of $dW/dz$ is then [14]

$$dW/dz \sim l_f^{-1} W_{1-\text{foil}} \sim 3/8 \alpha/b^2. \qquad (1)$$

We have assumed that the ordinary multi-foil interferences (not those due the rescattering effects discussed above) are washed out when integating over $\omega$ and $\theta$. We have $W_{1-\text{foil}} = 3/8 \alpha \gamma/b$. Eq.(1) suggests the existence of a universal bound of the form

$$dW/dz \leq C\alpha/b^2 \qquad (2)$$

for the energy emitted by any kind of long radiator at impact parameter $b$ [14].

## 3. Coherent radiation of the 6-MeV electron beam as a tool of shadowing effect investigation in macroscopic mode

The half-naked electron effect was considered in microscopic mode [1,2] for the case of the electron scattering on an atomic or nuclear structure. For the experimental confirmation of this effect the authors cite as evidence such experimental effects as Landau-Pomeranchuk effect [3] and density effect of Ter-Mikaelian [4], which may be used for half-naked electron concept confirmation only indirectly. However, the direct observation and investigation of this effect is not carried out yet.

The analysis of this phenomenon for the condition of the electron beam of microtron at Tomsk Nuclear Physics Institute allowed us to hope for a possibility of the investigation of this effect in a macroscopic mode. Actually shadowing of a relativistic electron electromagnetic field in conceiving of pseudo-photons may be achieved by pseudo-photon reflection in the mirror or by absorption in an absorber.

The effective transversal size of an electromagnetic field of a relativistic electron in respect to the transition or diffraction radiation generation is about $\gamma\lambda$, where $\gamma$ is the electron Lorenz-factor and $\lambda$ is a measured radiation wavelength. It is clear from the form of modified Bessel function argument in Fourier transform of the electron electromagnetic field [4,5]

$$K_1\left(\frac{2\pi\rho}{\gamma\lambda\cdot\beta}\right),$$

where $\rho$ is the transversal distance from electron trajectory to a considered point, β is an electron velocity in units of light one. For instance for $\gamma=12$ and $\lambda=10$ mm the effective transversal size of an electron electromagnetic field is approximately equal to 12 cm. This is quite a macroscopic size. Furthermore, the distance of the electron electromagnetic field recovering, proposed in [1] and [2] as $\gamma^2\lambda$, for shown conditions is equal $\approx 1.4$ m.

One may argue that the radiation intensity in this wavelength region during the interaction of the electron field with the targets is negligible and not accessible for a measurement. However, this is not the case because of the coherent character of radiation. Actually if the number of electron in a bunch $N_e \gg 1$, the radiation intensity from the bunch may be presented as

$$I_b = N_e \cdot \left(1 + N_e \cdot f^2\left(\frac{\lambda}{\sigma}\right)\right) \cdot I_e,$$

where $f\left(\frac{\lambda}{\sigma}\right)$ is a form-factor of electron bunch, $\sigma$ is the r.m.s. bunch length.

For $\lambda \gg \sigma$   $f\left(\frac{\lambda}{\sigma}\right) \approx 1$ and $I_b \approx N_e^2 \cdot I_e$ instead $I_b \approx N_e \cdot I_e$, which take place for incoherent radiation. For the electron beam of microtron at Tomsk Nuclear Physics Institute with the following parameters

- The electron energy was equal to 6.1 MeV ($\gamma = 12$).
- The macro-pulse of accelerated electrons with the duration of $\tau \approx 4$ $\mu$sec consisted of a train of $n_b \approx 1.6\cdot 10^4$ bunches with a period of 380 psec.
- The bunch population is equal to $N_e=6\cdot 10^8$.
- The bunch length was measured by using a coherent Smith-Purcell radiation technique [6]. From the approximation of the bunch longitudinal distribution by a Gaussian we have determined the length parameter σ≈1.6mm.

the coherent radiation intensity for λ>9 mm is by 8 orders larger than incoherent one and has the power level ≈1 Watt per steradian. It means one may investigate coherent radiation in this wavelength range without a problem.

As it was mentioned above, the shadowing of a relativistic electron electromagnetic field may be achieved by pseudo-photon reflection by the mirror or by absorption in absorber. The substantiation of this method is based on the fact that for γ>>1 the electron field properties became close to the properties of an electromagnetic field in free space [5]. Pseudo-photons are reflected in the mirror and are absorbed in absorber almost as well as real photons.

Following this assumption a diffraction radiation target may be used as an analyzer of the electron field downstream to the shadowing source. This procedure may be realized by the measurement of backward diffraction radiation (BDR) angular distribution according the scheme, shown in Fig. 7.

It does not matter whether we use an absorber or mirror for shadowing, but the application of an absorber is more conclusive.

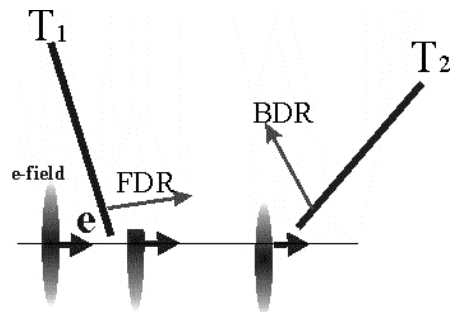

Fig. 7. Principle scheme for electron field shadowing investigation. $T_1$ is an absorber or conductive target, $T_2$ is a conductive mirror.

## 4. Theoretical calculation

For the calculation of expected angular distribution of the radiation from a conductive target $T_2$ (see Fig. 7) we shall use the above specified statement that the shadowed electron field may be presented as a sum of the electron field in free space and the forward diffraction radiation (FDR) emitted from absorber. The BDR angular distribution of the shadowed electron electromagnetic field from the conductive target $T_2$ (Fig. 7) may be presented as an interference of the FDR from the target $T_1$ reflected by the target $T_2$, and BDR from a conductive target $T_2$.

### 4.1. *Bound and released electron fields*

The results presented below are derived in the ultrarelativistic approximation. The Coulomb field of an electron fast moving along the $z$-axis is[†]

$$\mathbf{E}_C(t,x,y,z) \simeq \frac{-e}{4\pi} \gamma \left[ \mathbf{r}_T^2 + (z-vt)^2 \right]^{-3/2} \mathbf{r}_T \qquad (3)$$

We have neglected the longitudinal component. Using partial Fourier transformation, it can also be represented in energy-position space,

$$\mathbf{E}_C(\omega,\mathbf{r}) = \frac{-e\,\omega\,\mathbf{r}_T}{2\pi\gamma v^2 \,|\mathbf{r}_T|} K_1\left( \frac{\omega\,|\mathbf{r}_T|}{\gamma v} \right) e^{i\omega z/v}, \qquad (4)$$

or in energy, transverse momentum and $z$ space,

$$\mathbf{E}_C(\omega,\mathbf{k}_T,z) = \frac{ie}{v} \frac{\mathbf{k}_T}{\mathbf{k}_T^2 + q_0^2} e^{i\omega z/v}, \qquad (5)$$

or in energy, $x, k_y$ and $z$ space,

$$\mathbf{E}_C(\omega,x,k_y,z) = \frac{ie}{2v} e^{-\mu|x|} e^{i\omega z/v} (i\,\text{sign}(x),\tau), \qquad (6)$$

with $q_0 = \omega/(\gamma v)$, $\mu = \sqrt{k_y^2 + q_0^2}$, $\tau = k_y/\mu$.

The Coulomb field is made of virtual, or bound photons, of phase velocity $\omega/k_z = v$, whereas diffraction radiation is made of real photons, i.e., $\omega/|\mathbf{k}| = 1$. If the electron is suddenly stopped, or scattered at large angle, at $z = z_0$ then the field (3-6) is "released" and becomes the free field $\mathbf{E}_{rl}$. Its evolution along the $z$-axis is given by

$$\mathbf{E}_{rl}(z) = G(z - z_0) \mathbf{E}_C(z_0), \qquad (7)$$

where $G$ is the free propagator from transverse plane $z_0$ to transverse plane $z$. In the $\mathbf{r}_T$ representation, $G(z)$ is the convolution factor

---

[†] We use relativistic units ($c$=1) and rational definitions of fields and charge, for instance $\nabla \mathbf{E} = \rho$, $e^2/(4\pi\hbar)$=1/137.

$$G(\mathbf{r}_T; z) = (iz\lambda)^{-1} \exp\left\{i\omega\left(z + \frac{\mathbf{r}_T^2}{2z}\right)\right\}, \tag{8}$$

which is the paraxial approximation of the Huyghens-Fresnel formula. In the $\mathbf{k}_T$ representation, $G$ is multiplication factor:

$$G(\mathbf{k}_T; z) = \exp\left\{i\left(\omega - \frac{k_T^2}{2\omega}\right)z\right\}. \tag{9}$$

In the $(x, k_y)$ representation

$$G(x, k_y; z) = (iz\lambda)^{-1/2} \exp\left\{i\omega z + i\omega\frac{x^2}{2z} - i\frac{k_y^2}{2\omega}z\right\}. \tag{10}$$

For instance, (5) becomes

$$\mathbf{E}_{rl}(\omega, \mathbf{k}_T, z) = \frac{ie}{v}\frac{\mathbf{k}_T}{\mathbf{k}_T^2 + q_0^2} \exp\left\{i\omega\frac{z_0}{v} + i\left(\omega - \frac{k_T^2}{2\omega}\right)(z - z_0)\right\}. \tag{11}$$

The $z$ evolution relates the $\mathbf{r}_T$ representation at large $Z = z - z_0$ to $\mathbf{k}_T$ representation at $z_0$ by

$$\mathbf{E}_{rl}(\omega, \mathbf{r}_T, z_0 + Z) \sim \frac{\omega}{2\pi Z}\exp\left\{i\omega\left(Z + \frac{r_T^2}{2Z}\right)\right\}\mathbf{E}_{rl}(\omega, \mathbf{k}_T, z_0)|_{\mathbf{k}_T = \omega \mathbf{r}_T / Z}. \tag{12}$$

At small angle $\vec{\theta} \simeq \mathbf{r}_T/Z$ and for large $\gamma$ we recover the result derived from the retarded potential of the stopping electron,

$$E_{ret} \simeq e^{iz_0/v}\frac{e}{2\pi R}e^{i\omega R}\frac{\vec{\theta}}{\theta^2 + \gamma^{-2}}, \tag{13}$$

with $\mathbf{R} = (\mathbf{r}_T, z - z_0)$. This shows the validity of the "released photon" picture for the radiation at small angle by large $\gamma$ electrons.

### 4.2. *Diffraction Radiation from one target*

Let the diffraction radiation target $T_1$ be in the $z = z_1$ plane. The forward diffraction radiation field $\mathbf{E}_{FDR}$ can be considered as the negative of the intercepted Coulomb field, propagating freely in the forward half-space (Fig.8*a*) :

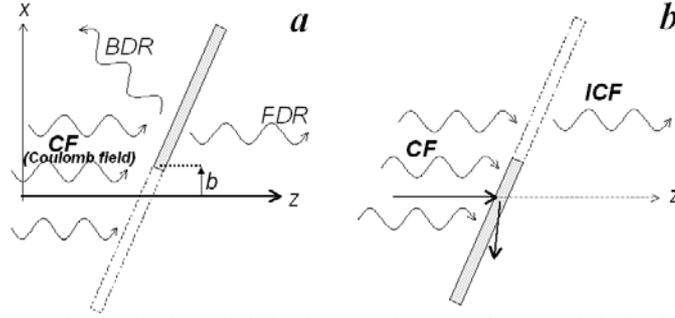

Fig. 8. Presentation of the forward diffraction radiation as an intercepted Coulomb field (ICF), propagating freely in the forward half-space.

$$\mathbf{E}_{FDR}(z) = -G(z-z_1) A \mathbf{E}_C(z_1) \qquad (14)$$

where $A(x, y)$ is the supporting function of the transverse area of the target. In our case, $A(x, y) = \theta(x - b)$. Forward transition radiation is obtained in the limit $b = -\infty$. $\mathbf{E}_{FDR}$ is independent on the target material (provided it is fully opaque), on the surface roughness and is unchanged when the screen is tilted. The field after the screen is the sum of FDR and the full Coulomb field $\mathbf{E}_C$ of (3-6). Just behind the screen they interfere destructively so that the total field is extremely small. Then the interference gradually disappears due to the different evolutions in $z$ (forced versus free) of the two fields.

For a mirror target, the backward diffraction radiation field $\mathbf{E}_{BDR}$ is symmetrical of $\mathbf{E}_{FDR}$ with respect to the target plane.

If the electron stops at $z_1$, a masked released field

$$\mathbf{E}_{mrl}(z) = +G(z - z_1) \overline{A} \mathbf{E}_C(z_1) \qquad (15)$$

occupies the forward region alone. $\overline{A}(x, y) = 1 - A(x, y)$ is the supporting function of the complementary target. The FDR field from a target T is the opposite of the masked released field from the complementary target $\overline{T}$ :

$$\mathbf{E}_{FDR}\{T\} = -\mathbf{E}_{mrl}\{\overline{T}\}. \qquad (16)$$

The FDR amplitude at $z = z_1$ is most conveniently calculated in the $(x, k_y)$ representation (6):

$$\mathbf{E}_{\text{FDR}}(\omega, x, k_y, z_1) = -\frac{ie}{2v}\theta(x-b)e^{-\mu|x|}e^{i\omega z_1/v}(i\,\text{sign}(x), \tau) \quad (17)$$

We have approximated $v$ by 1, except in the exponentials. Taking the Fourier transform in $k_x$, we obtain the full $\mathbf{k}_T$ representation which, for $b > 0$ (non-intercepting screen), reads

$$\mathbf{E}_{\text{FDR}}(\omega, \mathbf{k}_T, z_1) = -\frac{ie}{2v(\mu + ik_x)}e^{-(\mu + ik_x)b}e^{i\omega z_1/v}(i, \tau) \quad (18)$$

Applying (12-13) to $\mathbf{E}_{\text{FDR}}$ one obtains the well-known energy-angle spectrum of Diffraction Radiation (here $\mathbf{k}_T = \omega\vec{\theta}$):

$$\frac{dW}{\hbar d\omega d\Omega} = I(\omega, \vec{\theta}) = \frac{R^2}{\pi}|\mathbf{E}_{\text{FDR}}(\omega, \mathbf{R})|^2 = \frac{\alpha\omega^2}{4\pi^2\mu^2}\frac{q_0^2 + 2k_y^2}{q_0^2 + \mathbf{k}_T^2}e^{-2\mu b}. \quad (19)$$

### 4.3. *Generalization to two targets*

Let us now consider two successive targets $T_1$ and $T_2$, at abscissa $z_1$ and $z_2$ and supporting functions $A_1$ and $A_2$. The field between $T_1$ and $T_2$ is

$$\mathbf{E}_{\text{betw}}(z) = \mathbf{E}_{\text{FDR}}^{(1)}(z) + \mathbf{E}_C(z), \quad (20)$$

with $\mathbf{E}_{\text{FDR}}^{(1)}(z)$ given by (14). In the backward direction from the mirror $T_2$ one observes the reflection of $\mathbf{E}_{\text{betw}}$ by $T_2$:

$$\mathbf{E}_{\text{back2}} = -S_2\{G(z-z_2)A_2\mathbf{E}_{\text{betw}}(z_2)\} \quad (21)$$

where $S_2$ is symmetry operator about the $T_2$ plane. The energy-angle distribution of $\mathbf{E}_{\text{back2}}$ is determined by the $\mathbf{k}_T$ representation of the field

$$\mathbf{F} = A_2\mathbf{E}_{\text{int}}(z_2) = A_2 G(z_2 - z_1)\mathbf{E}_{\text{FDR}}^{(1)}(z_1) - \mathbf{E}_{\text{FDR}}^{(2)}(z_2) \equiv \mathbf{F}_1 - \mathbf{F}_2, \quad (22)$$

where (14) has been used for target $T_2$.
The analytical calculations of expression for **F** are presented in **Appendix A**.

### 4.4. *Numerical calculations*

Numerical calculations of the radiation angular density *W* were performed for shown above electron beam parameters, for e=1, impact-parameter h=10 mm and wavelength λ=10 mm. First we consider the calculations for case where targets are on the same side of electron beam (see Fig. 7). In Fig. 9 the radiation angular distribution for different values of the distance L between absorber and conductive target is shown. Fig. 10 shows the same dependence for value interval of L= 0~220 mm, which corresponds to the experimental conditions.

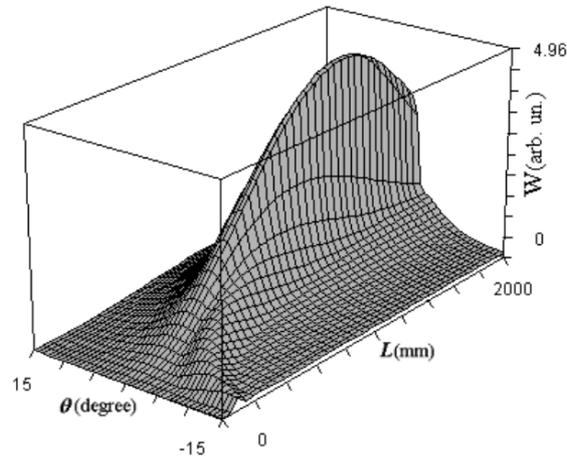

Fig. 9. Calculated dependence of the radiation intensity on the observation angle and distance to the absorber according to the scheme shown in Fig. 7.

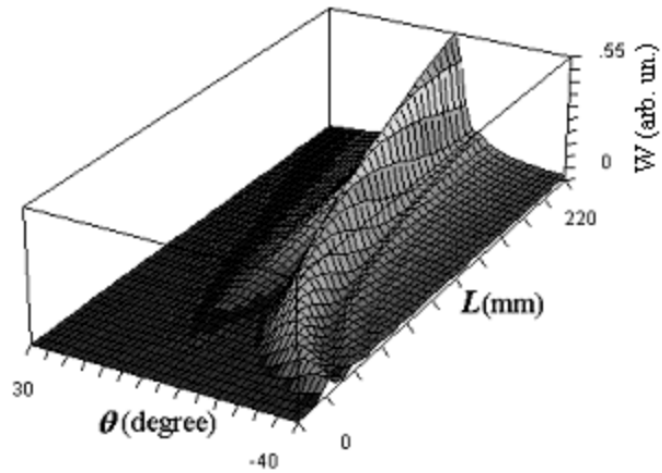

Fig. 10. Dependence shown in Fig. 9 in the range of the experimental conditions.

In figures 9 and 10 we can see the manifestation of the shadowing effect and the effect of the recovering of the electron field.

For the case where target are on opposite sides of the beam we keep the same $T_1$ target, with $b_1 > 0$, but put $T_2$ on the opposite side (see Fig. 11).

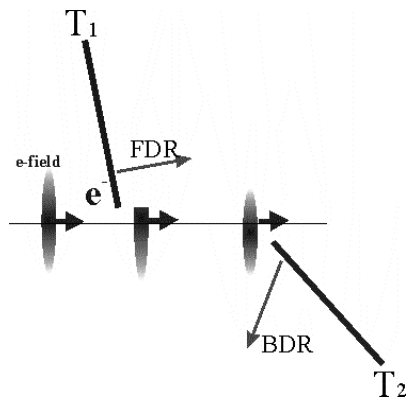

Fig. 11. The scheme for calculation of the radiation for the case of opposite position of the absorber.

The result of numerical calculations for this geometry is shown in Fig. 12.

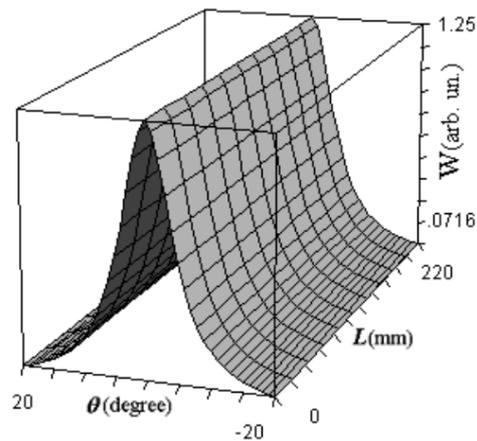

Fig. 12. Calculated dependence of the radiation intensity on the observation angle and distance between the targets in case of opposite position of the targets

It is seen on Fig. 12, that in this case the influence of the shadowing is negligible.

## 5. Experimental setup and method

The experiment was performed on the extracted electron beam of microtron of Tomsk Nuclear Physics Institute with parameters, shown above. The electron beam is extracted from the vacuum chamber through the beryllium foil with the thickness of 20 μm. The beam divergence caused by beryllium window (≈0.08 radian) limits the achievable distance between the absorber and a conductive target by the value $L_{max}$≈220 mm because of the necessity of impact parameter increase for providing the diffraction radiation mode. To exclude the transversal bunch form-factor contribution the position of the conductive target was fixed and the distance $L$ was changed by the varying of the absorber position.

The beryllium window with diameter of 34 mm may be also considered as a source of a transition radiation. The simple calculations show that the intensity of this radiation downstream to the absorber is negligible.

For the radiation measurements we used the room temperature detector DP20M, with parameters described in [11]. Main elements of the detector are the low-threshold diode, broadband antenna and preamplifier. The detector efficiency in the wavelength region $\lambda$=3~16 mm is estimated as a constant with accuracy ± 15%. The detector sensitivity is 0.3 V/mWatt. The beyond-cutoff wave-guide with $\lambda_{cut}$=17 mm was used to cut the accelerator RF system long wave background. The high frequency limit of wavelength interval is defined by bunch form-factor. This limit ($\lambda_{min}$ ≈9 mm) was measured using discrete wave filters [7] and the spectrometer type of grating.

To exclude the prewave zone effect contribution (see [8]) the parabolic telescope was used for an angular distribution measurement. This method was suggested and tested in [9] and allows us to measure an angular distribution coinciding with one in far field zone ($R >> \gamma^2\lambda$).

The absorber properties were checked both on the real photon beam from the radiation source with wavelength $\lambda$=6 mm and by the measurement of reflected pseudo-photons of beam electrons according to the scheme shown in Fig. 13.

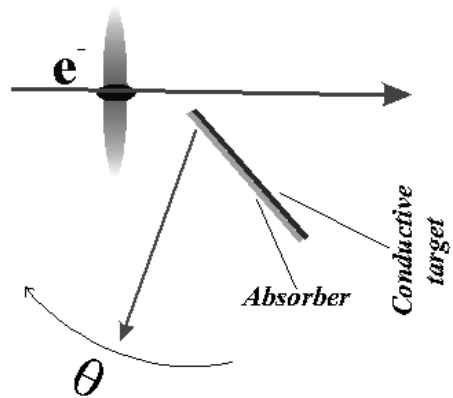

Fig. 13. Scheme for the measurements of the absorber properties.

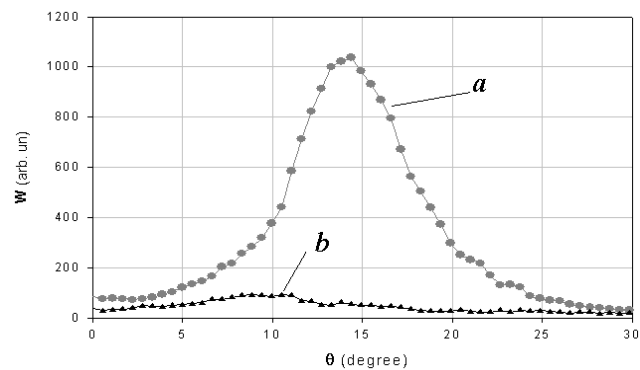

Fig. 14. Angular distribution of the radiation intensity, measured using scheme shown in Fig. 13.
*a* – BDR from conductive target without absorber
b – Radiation from the same target covered by absorber

In Fig. 14 curve *a* is the BDR angular distribution from conductive target without absorber and curve *b* is the similar distribution from conductive target covered by absorber. We can see that almost no pseudo-photon reflection is registered. The test of the absorber on the real photon beam with accuracy of the experiment error ($\approx 3\%$) showed that no real photons passed through absorber and no photons were reflected by it.

## 6. Experimental results

Using the described method the angular distributions of backward radiation from a conductive target for different distance *L* between absorber and a conductive target were measured for $L$=0 to 220 mm with the step of 20 mm. Two schemes of the experiment were used for measurements.

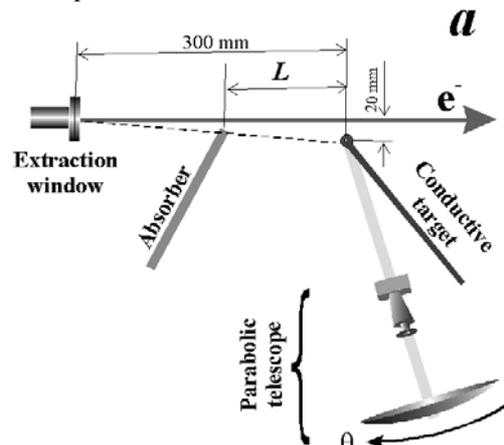

Fig. 15. Scheme of experiment on the shadow effect observation.

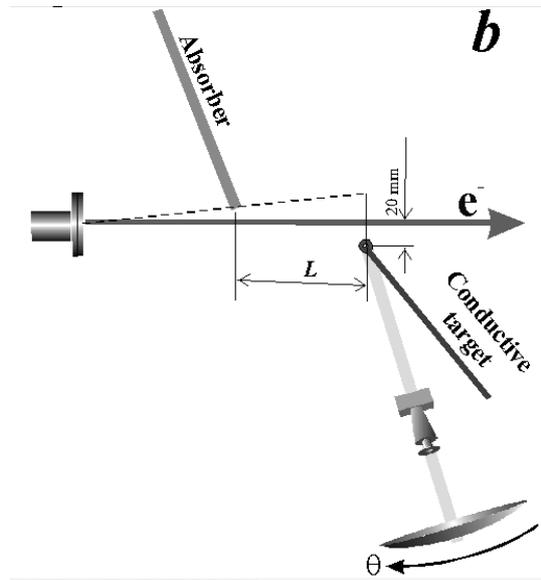

Fig. 16. Scheme of experiments with opposite position of absorber.

Scheme *a* (Fig. 15) is intended for demonstration of the shadowing of electron field. To extend the measured angular range of radiation from a conductive target for small value of *L* the absorber was inclined at the angle of $45^0$ as it is shown in Fig. 15.

In scheme ***b*** (Fig 16) the absorber is placed on the opposite side to the electron beam. In this geometry the shadow effect is expected not to be observed.

The samples of the measured distributions using scheme ***a*** are shown in Fig. 17 and the total dependence on the observation angle and on the distance *L* is shown in Fig. 19***a***.

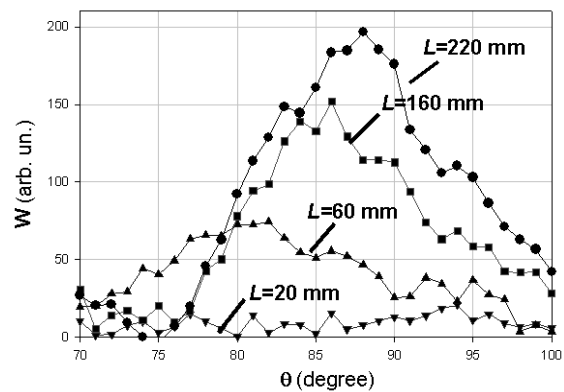

Fig. 17. The samples of the measured angular distributions for different distance to the absorber using scheme ***a*** (Fig. 15)

Accordingly in figures 18 and 19***b*** the samples of angular distributions and the total dependence, measured using scheme ***b*** are presented.

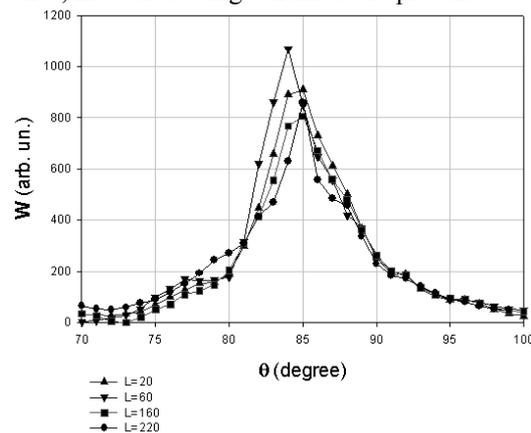

Fig. 18. The samples of the measured angular distributions for different distance to the absorber using scheme ***b*** (Fig. 16)

Both Fig. 17 and Fig 18 have the same scale of radiation intensity.

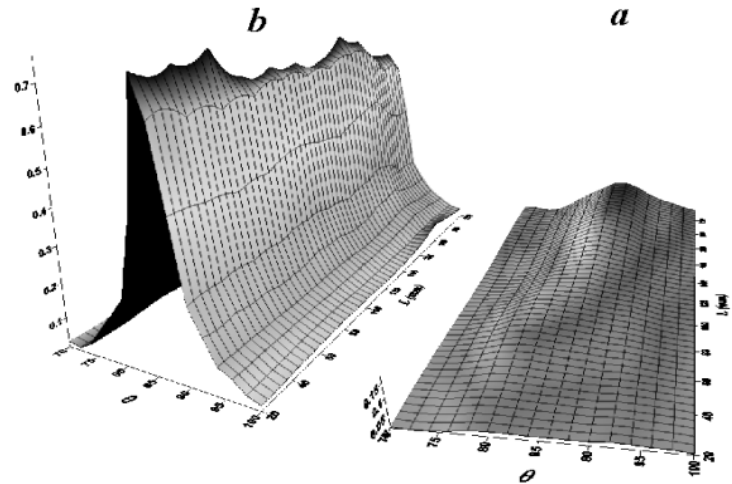

Fig. 19. Measured dependence of the radiation intensity on the observation angle and distance to the absorber
*a* – measured using scheme of experiment, shown in Fig. 15.
*b* – measured using scheme of experiment, shown in Fig. 16 with opposite position of the absorber (no shadow effect).

Figures 19*a* and 19*b* are presented in the same scale of radiation intensity to emphasize the shadowing effect.

## 7. Discussion

The present experiment has demonstrated the feasibility of a direct observation of the shadowing of an electron electromagnetic field in a macroscopic mode. Unfortunately the experimental conditions in this experiment could not allow us to measure the *L*-dependence of the recovering up to $\gamma^2\lambda$ (*L* is the distance from the absorber). Nevertheless, at the measured distances, the dependence of the radiation intensity on *L* and on the observation angle is in a good agreement with the theoretical one.

The used experimental method allowed us to make a basic analysis of this phenomenon. It is mentioned above that the shadowed electron field may be presented as a sum of the electron field in free space and the FDR emitted from the absorber (FTR if the electron crosses the absorber). This statement may lead to some confusion concerning the nature of the radiation. Actually in a traditional interpretation, transition and diffraction radiation from perfectly conducting targets are considered as radiations emitted by surface currents (for example in [10]). More generally the DR or TR source is treated in macroscopic electrodynamics as a polarization current, expressed through the dielectric permeability. These surface or polarization currents are induced by the electromagnetic field of the relativistic electron.

However, as it is mentioned above, for $\gamma \gg 1$ the electron field properties become close to those of an electromagnetic field in free space, and pseudo-photons are reflected by the mirror or absorbed in the absorber almost like real photons. In the absorber case the electron electromagnetic field has disappeared on the downstream side and no surface current or polarization current may occur on this side (the shadowing effect). One could then conclude (in this traditional interpretation) that no forward transition or diffraction radiation may be generated from an absorber. The mistake in this reasoning is : either one uses Maxwell equations in vacuum (which do not include absorption) and consider the polarizations currents as field sources, or one uses Maxwell equations in matter (which include absorption) but do not take the polarizations currents as sources. It is therefore not correct to "screen" the field of the polarization currents. The same can be said for a perfectly conducting thin target, which is a limiting case.

Alternatively, one may attribute forward optical transition radiation (FTR) to the current of the traveling electron after traversal of the screen. Indeed this radiation is practically the same as that of a suddenly accelerated electron. At the same time this current re-creates the Coulomb field. We can then say that FTR is generated during the transit of electron from the unstable "naked" state to the stable "dressed" state (here the term "transition radiation" seems particularly appropriate). We can find a similar interpretation of this phenomenon in [1]. As for forward DR, it can be viewed as a consequence of the transit from a "half-naked" state to the "dressed" state.

The two points of view (polarization currents and naked-dressed transition) are in fact dual. Calling $F_u$ the (retarded) field from the upstream electron trajectory (suddenly stopped electron) $F_d$ the field from the downstream trajectory (suddenly accelerated electron), $F_c$ the Coulomb field and $F_m$ the field emitted by the induced polarization currents, one has in the forward region

$$F_{\text{FTR}} = F_m = -F_u, \quad F_C = F_u + F_d \longrightarrow F_{\text{FTR}} = F_d - F_C.$$

Far enough from the trajectory, $F_{FTR}=F_d$ (second point of view). $F_{FTR}=F_m$ is the first point of view. $F_{FTR}=-F_u$ explains why the spectral-angular of backward and forward transition (or diffraction) radiations coincide. It remains that the second point of view is more intuitive for FTR [15].

**Appendix A. Analytical calculations.**

**A.1.** *Targets on the same side.*

We will work in the $(x, k_y)$ representation. Let us calculate the first contribution $\mathbf{F}_1$ to (22). We assume $b_1, b_2 > 0$ and take $z_1 = 0$, $z_2 = L$. Using (10) and (17),

$$\mathbf{F}_1(\omega, x_2, k_y, z_2) = -\frac{ie}{2} \theta(x - b_2)(i, \tau) \exp\left\{i\left(\omega - \frac{k_y^2}{2\omega}\right)L\right\}$$

$$\times (i\lambda L)^{-1/2} \int_{b_1}^{\infty} dx_1\, e^{-\mu x_1} \exp\left\{i\omega \frac{(x_2 - x_1)^2}{2L}\right\} \quad (23)$$

In full $\mathbf{k}_T$ space,

$$\mathbf{F}_1(\omega, \mathbf{k}_T, z_2) = -\frac{ie}{2}(i, \tau) \exp\left\{i\left(\omega - \frac{k_y^2}{2\omega}\right)L\right\} \times (i\lambda L)^{-1/2}$$

$$\int_{b_1}^{\infty} dx_1 \int_{b_2}^{\infty} dx_2 \exp\left\{-\mu x_1 - ik_x x_2 + i\omega \frac{(x_2 - x_1)^2}{2L}\right\} \quad (24)$$

To evaluate the double integral $J$, we divide the integration domain in two parts : in the region $x_2 - b_2 \geq x_1 - b_1 \geq 0$, we take $x_1$ and $u = x_2 - x_1$ as integration variables. In the region $x_1 - b_1 \geq x_2 - b_2 \geq 0$, we take $x_2$ and $v = x_1 - x_2$ :

$$J = \int_{b_1}^{\infty} dx_1 \exp\{-\mu x_1 - ik_x x_1\} \int_{b_2 - b_1}^{\infty} du \exp\left\{-ik_x u + i\omega \frac{u^2}{2L}\right\}$$

$$+\int_{b_2}^{\infty} dx_2 \exp\{-\mu x_2 - ik_x x_2\} \int_{b_1-b_2}^{\infty} dv \exp\left\{-\mu v + i\omega \frac{v^2}{2L}\right\}. \quad (25)$$

Thus $J = J_1 J_u + J_2 J_v$, where
$$J_i = (\mu + ik_x)^{-1} \exp[-(\mu + ik_x)b_i], \quad i = 1, 2$$
are the $x_1$ and $x_2$ integrals, and $J_u$, $J_v$ are the $u$- and $v$- integrals. The $J_u$ path can be rotated by the angle $+\pi/4$ in the complex plane, with the change of variable $t = (2iL/\omega)^{-1/2}(u - Lk_x/\omega)$. It yields

$$J_u = (2i\pi L/\omega)^{1/2} \exp\left(-i\frac{Lk_x^2}{2\omega}\right) \frac{1}{2} \text{Erfc}\left[(2iL/\omega)^{-1/2}(b_2 - b_1 - k_x L/\omega)\right] (26)$$

where $\text{Erfc}(z) = 2\pi^{-1/2} \int_z^{+\infty} dt\, e^{-t^2}$ is the error function for complex argument. For the $v$-integral we interchange $b_1$ and $b_2$ and replace $k_x$ by $-i\mu$:

$$J_v = (2i\pi L/\omega)^{1/2} \exp\left(+i\frac{L\mu^2}{2\omega}\right) \frac{1}{2} \text{Erfc}\left[(2iL/\omega)^{-1/2}(b_1 - b_2 + i\mu L/\omega)\right] (27)$$

Gathering (10), (17), (18), (22) and (24-27),

$$\mathbf{F}(\omega, \mathbf{k}_T) = \frac{ie}{2}(i,\tau) \frac{e^{-(\mu + ik_x)b_2}}{\mu + ik_x} \{e^{i\omega L/v} - e^{i\omega L}[$$

$$e^{(\mu + ik_x)(b_2 - b_1)} \exp\left(-iL\frac{k_T^2}{2\omega}\right) \frac{1}{2} \text{Erfc}\left[(2iL/\omega)^{-1/2}(b_2 - b_1 - k_x L/\omega)\right]$$

$$+ \exp\left(iL\frac{\mu^2 - k_y^2}{2\omega}\right) \frac{1}{2} \text{Erfc}\left[(2iL/\omega)^{-1/2}(b_1 - b_2 + i\mu L/\omega)\right]]\} \quad (28)$$

One can check that $\mathbf{F}(\omega,\mathbf{k}_T)$ is zero for $L \simeq 0$ and $b_1 < b_2$ (full shadowing) and is equal to $\mathbf{E}_{FDR}(b_2) - \mathbf{E}_{FDR}(b_1)$ (partial shadowing) for $L \simeq 0$ and $b_1 > b_2$.

### A.2. *Targets on opposite sides.*

Here we keep the same $T_1$ target, with $b_1 > 0$, but put $T_2$ on the opposite side, with transverse supporting function $A_2(x,y) = \theta(b_2 - x)$, $b_2 < 0$. The relevant field is again given by (22), *i.e.*,

$$\mathbf{F}_{opp} = A_2\, G(z_2 - z_1)\mathbf{E}_{FDR}^{(1)}(z_1) - \mathbf{E}_{FDR}^{(2)}(z_2), \qquad (29)$$

Considering the screen $\overline{T}_2$ complementary to $T_2$, we can rewrite it

$$\mathbf{F}_{opp} = (1 - \overline{A}_2)\, G(z_2 - z_1)\mathbf{E}_{FDR}^{(1)}(z_1) - \mathbf{E}_{FDR}^{(2)}(z_2)$$

$$= \mathbf{E}_{FDR}^{(1)}(z_2) - \mathbf{E}_{FDR}^{(2)}(z_2) - \mathbf{F}_1\{T_1, \overline{T}_2\}, \qquad (30)$$

$\mathbf{F}_1\{T_1, \overline{T}_2\} = \mathbf{F}_1(b_1, b_2,...)$ being given by the same expression as for the same-side case (see Eq.(9) without the term in $e^{i\omega L/v}$).

From (18) and (9), we have

$$\mathbf{E}_{FDR}^{(1)}(\omega, \mathbf{k}_T, z_2) = -\frac{ie(i,\tau)}{2(\mu + ik_x)} e^{-(\mu + ik_x)b_1} \exp\left\{i\omega z_1/v - iL\frac{k_T^2}{2\omega}\right\} \quad (31)$$

$$\mathbf{E}_{FDR}^{(2)}(\omega, \mathbf{k}_T, z_2) = -\frac{ie(-i,\tau)}{2(\mu - ik_x)} e^{(\mu - ik_x)b_2} \exp(i\omega z_2/v), \qquad (32)$$

Gathering (31), (32) and (28) without the term in $e^{i\omega L/v}$, we obtain

$$\frac{2}{ie}\mathbf{F}_{opp}(\omega, \mathbf{k}_T) = \frac{(-i,\tau)}{\mu - ik_x} e^{(\mu - ik_x)b_2} \exp(i\omega L/v) - \frac{(i,\tau)}{\mu + ik_x} e^{i\omega L} \{$$

$$e^{-(\mu + ik_x)b_1} \exp\left(-iL\frac{k_T^2}{2\omega}\right)\frac{1}{2}\mathrm{Erfc}\left[(2iL/\omega)^{-1/2}(b_2 - b_1 - k_x L/\omega)\right]$$

$$+e^{-(\mu+ik_x)b_2}\exp\left(iL\frac{\mu^2-k_y^2}{2\omega}\right)\frac{1}{2}\text{Erfc}\left[(2iL/\omega)^{-1/2}(b_1-b_2+i\mu L/\omega)\right]\}.$$
(33)